\documentclass[a4paper]{article}

\usepackage[english]{babel}
\usepackage[utf8]{inputenc}
\usepackage{csquotes}
\usepackage{hyperref}
\usepackage{amsmath}
\usepackage{stmaryrd}
\usepackage{amssymb}
\usepackage{graphicx}
\usepackage{subcaption}
\usepackage{layaureo}
\hypersetup{colorlinks=true,citecolor=blue,urlcolor=blue,linkcolor=blue}
\usepackage[backend=biber, style=phys, eprint=true, hyperref=true, biblabel=brackets, url=true]{biblatex}
\usepackage{cleveref}
\DeclareFieldFormat*{title}{\textit{#1}} 
\DeclareSourcemap{
  \maps[datatype=bibtex, overwrite=true]{
    \map{
      \step[fieldset= eprintclass, fieldvalue=] 
    }
    \map{
      \step[fieldsource=collaboration,final=true] 
      \step[fieldset=usera,origfieldval,final=true]
    }
  }
}
\renewbibmacro*{author}{%
  \iffieldundef{usera}{%
    \printnames{author} 
  }{%
    \printfield{usera}, \printnames{author} 
  }
}
\DeclareFieldFormat{eprint}{%
  \ifhyperref
    {\href{https://arxiv.org/abs/#1}{\texttt{arXiv:#1}}}
    {\texttt{arXiv:#1}}
}
\DeclareCiteCommand{\citeurl}
  {\usebibmacro{prenote}}
  {\usebibmacro{citeindex}%
   \iffieldundef{eprint}
     {\printfield{url}}
      {\printfield{eprint}}
     }
  {\multicitedelim}
  {\usebibmacro{postnote}}

\title{Graph-based Full Event Interpretation: a graph neural network for event reconstruction in Belle II }
\author{M. AbuMusabh, J. Cerasoli, G. Dujany, C. Santos}

\begin{document}

\maketitle
\begin{center}
\textit{Proceedings of the 2024 Conference on Computing in High Energy and Nuclear Physics}
\end{center}

\abstract{In this work we present the Graph-based Full Event Interpretation
(GraFEI), a machine learning model based on graph neural networks to inclusively
reconstruct events in the Belle~II experiment.
Belle~II is well suited to perform measurements of $B$ meson decays involving
invisible particles (e.g. neutrinos) in the final state. The kinematical
properties of such particles can be deduced from the energy-momentum imbalance
obtained after reconstructing the companion $B$ meson produced in the event.
This task is performed by reconstructing it either from all the particles in an
event but the signal tracks, or using the Full Event Interpretation, an
algorithm based on Boosted Decision Trees and limited to specific, hard-coded
decay processes. A recent example involving the use of the aforementioned
techniques is the search for the $B^+ \to K^+ \nu \bar \nu$ decay, that provided
an evidence for this process at about 3 standard deviations.
The GraFEI model is trained to predict the structure of the decay chain by
exploiting the information from the detected final state particles only, without
making use of any prior assumptions about the underlying event. By retaining
only signal-like decay topologies, the model considerably reduces the amount of
background while keeping a relatively high signal efficiency. The performances
of the model when applied to the search for $B^+ \to K^+ \nu \bar \nu$ are
presented.  
}

\section{Introduction}

One of the main tasks in any particle physics analysis is the reconstruction of
decays. Since decays are often represented as decay trees and with the recent
advances in machine learning, Graph Neural Networks (GNNs) have become an
interesting tool for studying these decays.  Indeed, GNNs are a particular
class of neural networks acting on \textit{graphs} $\mathcal{G}$, entities
composed of $N \in \mathbb{N}^*$ nodes $V=\{v_i\}_{i\in \llbracket 1, N
\rrbracket}$, connected by edges
$E~=~\{e_{v_iv_j}~\equiv~e_{ij}\}_{(i,j)~\in~\llbracket 1,N \rrbracket^2, i\neq
j}$. In this perspective, we developed the Graph-based Full Event Interpretation
(\textsc{graFEI}), a machine learning model based on GNNs to inclusively
reconstruct events at the Belle~II experiment~\cite{abe2010belle}.

The Belle~II experiment analyses $e^+e^-$ collisions at the $\Upsilon(4S)$
resonance, producing $B\bar{B}$ pairs. Since the $B$ mesons are produced in
pairs, the kinematical properties of decays involving invisible particles
(\textit{e.g.} neutrinos) can be deduced from the energy-momentum imbalance
obtained after reconstructing the companion $B$ meson produced in the event.
This task is performed by reconstructing it either from all the particles in an
event but the signal tracks, called inclusive reconstruction, leading to a high
efficiency but a low signal purity, or using the \textsc{Full Event
Interpretation (\textsc{FEI})}~\cite{Full_Event_Recon}, an algorithm based on boosted
decision trees and limited to specific, hard-coded decay processes, giving a
result with a high signal purity but a low efficiency. A recent example
involving the use of the aforementioned techniques is the search for the $B^+
\to K^+ \nu \bar \nu$ decay, that provided an evidence for this process at about
3 standard deviations~\cite{Belle-II:2023esi}.

The goal of the \textsc{graFEI} model is to reconstruct the decay chain of the
companion $B$ meson produced in the event by exploiting the information from the
detected final state particles only, without making use of any prior assumptions
about the underlying event, as done in the inclusive reconstruction strategy.
By retaining only signal-like decay topologies, similarly to the \textsc{FEI}, the model considerably reduces
the amount of background while keeping a relatively high signal efficiency. The
performances of the model when applied to the search for $B^+ \to K^+ \nu \bar
\nu$ are presented in this work as a proof of concept.

\section{The \textsc{graFEI} model}
\subsection{Problem statement}\label{subsec:problem_statement}
Decay trees can be described by rooted directed acyclic tree graphs, represented
algebraically by matrices like the adjacency matrix, a binary matrix which rows
and columns are the nodes of the graph and its elements are either 0 or 1,
indicating if the two nodes are connected or not. This approach is not suitable
for reconstructing decay trees, because we would have to make
assumptions on the intermediate particles. The \textsc{graFEI} reconstructs
events inclusively using the information on the final state particles alone,
without any prior assumption about the structure of the decay, a technique shown
in~\cite{kahn_learning_2022, Tsaklidis:2122, reuter_lea_2022_22115} thanks to
the \textit{Lowest Common Ancestor (LCA)} matrix.  Each entry of this matrix
corresponds to a pair of final state particles, and its elements are the lowest
ancestors common to pairs of particles. To avoid the use of a unique identifier
for each ancestor, a system of classes is used: 6 for $\Upsilon (4S)$ mesons, 5
for $B$, 4 for $D^*$, 3 for $D$, 2 for $K_s^0$, 1 for $\pi^0$ or $J/\psi$ and 0
for particles not belonging to the decay tree.  This new representation of the
LCA is called \textit{LCAS} matrix, where the S stands for ``stage''. In
addition to the LCAS matrix, the mass hypotheses of the final state particles
can also be predicted. To do so, a second system of classes is defined: 1 for
$e$, 2 for $\mu$, 3 for $\pi$, 4 for $K$, 5 for $p$, 6 for $\gamma$ and 0 for
everything else.

Inside graphs information can be stored in nodes, edges, or the graph as a
whole. Thus $\mathcal{G}$ is characterized by three sets of features:
\begin{itemize}
    \item \textbf{Node features} The node features encode information about the final state particles. The model uses the following node features: 
        particle charge, $p_T$, $p_z$, d$r$, d$z$, electronID, protonID, muonID, kaonID, pionID, as well as some clusters' information. In the case where a value is not available (e.g. cluster variables computed for charged particles) the variable is set to 0.
    \item \textbf{Edge features} The edge features are quantities that depend on pairs of particles. 
        The model uses the following edge features: the cosine of the azimutal angle between particles'~momenta ($\cos{\theta}$) and the distance of closest approach between two tracks (\texttt{DOCA}). 
        In order to compute the \texttt{DOCA} for photons, the particle is assumed to pass through the origin of the coordinate system.
    \item \textbf{Global features} The global features can be information about the overall graph (e.g.\ the number of final state particles, the total momentum, ...). 
        The model currently does not use any input global feature.
\end{itemize}

In order to reconstruct the LCAS matrix, the input of the network is a
\textit{fully connected} graph, where each node (i.e.\ each final state
particle) is connected with all the others. The output of the model is a graph
with the same structure as the input but updated values of its features.

\subsection{Model architecture}
\begin{figure}[htb]
    \centering
    \includegraphics[width=0.7\textwidth]{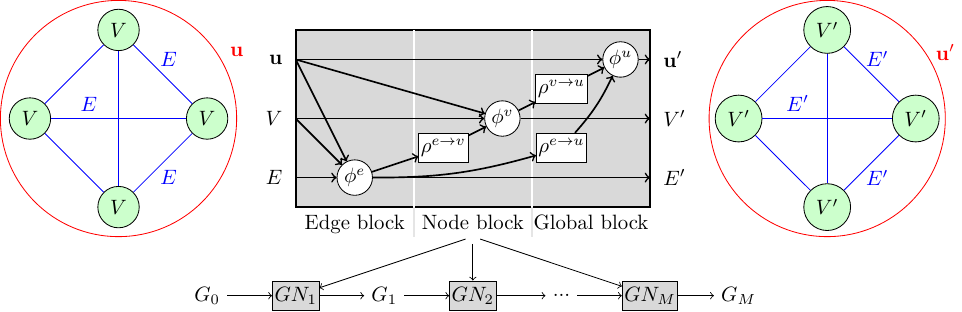}
    \caption{Schematic view of the \textsc{graFEI} model: the input graph $G_0$
    if passed through a series of graph network blocks ($GN_i$) each in turn
    composed of three sub-blocks. The output graph $G_M$ has the same structure
    of the input one but updated features, and is used to make predictions on
    the quantities of interest. Adapted
    from~\cite{battaglia2018}.}\label{fig:graFEI model}
\end{figure}

The \textsc{graFEI} model consists of a series of \textit{graph network
blocks}~\cite{battaglia2018}: the first block takes the input graph and updates
its features. The output graph is then passed through a series of intermediate
blocks (whose number is a hyperparameter of the model and is equal to 1 in the
default configuration), which further update the features. Finally, the last
block outputs a graph used to predict the quantities of interest. 
A \textit{skip-connection}~\cite{he2015deep} is present around each intermediate
graph network block.
A schematic view of the model is shown in Figure~\ref{fig:graFEI model}.
\\Each graph-network block is in turn composed of three sub-blocks: the edge, node, and global sub-blocks:
\begin{itemize}
    \item \textbf{Edge sub-block} The edge sub-block is used to update the edge
    features. The core of the block is a MultiLayer Perceptron (MLP) $\phi^e$,
    called the \textit{update function} of the sub-block.  For each edge $e$,
    the input vector of $\phi^e$ contains the edge features, the node features
    of the two nodes linked by the edge, and the global features. The output of
    the edge sub-block is also used as input to the node and global sub-blocks,
    via the use of \textit{aggregation functions} $\rho^{e \to v}$ and $\rho^{e
    \to u}$. For each output feature, these functions perform the arithmetic
    average for edges connected to the same node and for all the edges in the
    same graph, respectively. The output of the edge sub-block in the last
    graph network block is used to predict the LCAS matrix of the graph. This is
    done by evaluating probabilities for each edge to belong to the 6 classes
    defined in the LCAS matrix.
    \item \textbf{Node sub-block} The node sub-block is used to update the node
    features. The core of the block is a MLP $\phi^v$. For each node $v$, the
    input vector of $\phi^v$ contains the node features, the output of $\rho^{e
    \to v}$, and the global features. The output of the node sub-block is also
    used as input to the global sub-block, via the use of the aggregation
    function $\rho^{v \to u}$, which, for each output feature, performs the
    arithmetic mean for all the nodes in the same graph.  The output of the node
    sub-block in the last graph network block is used to predict the mass
    hypotheses of the final state particles. This is done by evaluating
    probabilities for each node to belong to the 7 classes defined for the mass
    hypothesis.
    \item \textbf{Global sub-block} The global sub-block is not used to predict
    physical quantities for the time being, nonetheless it is used to propagate
    the information throughout the network. The core of the block is a MLP
    $\phi^u$. The input vector of $\phi^u$ contains the global features and the
    output of $\rho^{e \to u}$ and $\rho^{v \to u}$. 
  \end{itemize}

The number of hidden layers of each MLP $\phi$ is a hyperparameter of the model.
In the default configuration it is equal to 2 and is the same for all the
update functions.

This model is implemented using the \textsc{PyTorch Geometric} library
\cite{fey2019fastgraphrepresentationlearning}, which provides a set of tools to
work with graph data structures and to implement graph neural networks.

\subsection{Training and evaluation}
The activation function used in the MLP is the Exponential Linear Unit~\cite{clevert2016fast}.

During the training phase, \textit{dropout}~\cite{hinton2012improving} is used
in order to reduce overtraining. It consists in
randomly setting to 0 the output of some neurons. The probability is an
hyperparameter of the model which is chosen to be $0.3$ in the default setting.
Dropout is applied to the first and hidden layers of each MLP, but not to the
output layer.

In order to accelerate the training of the network, \textit{batch
normalization}~\cite{ioffe2015batch} is applied to the output layer of each
MLP\@. Batch normalization is not applied in the last graph network block. 

\begin{figure}[htb]
    \centering
    \includegraphics[width=\textwidth]{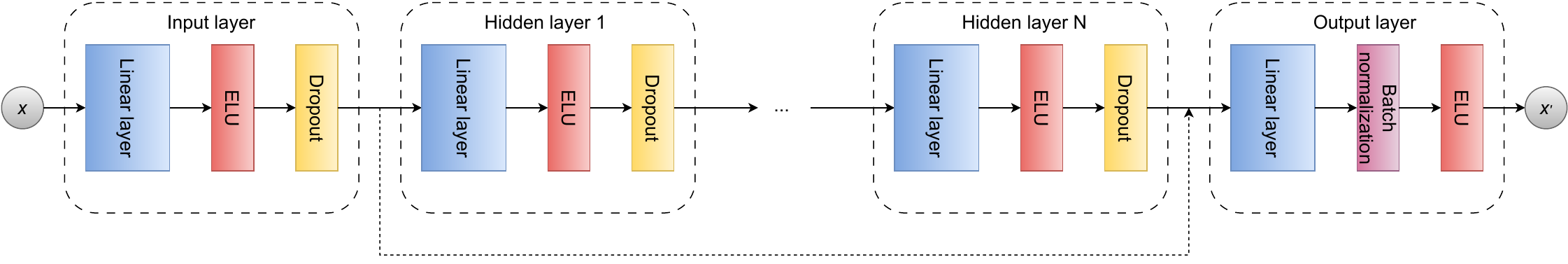}
    \caption{Schematic view of the structure of a MLP $\phi$ in the model. Batch normalization is not applied in the last graph network block. The dashed arrow indicates a skip-connection, which is present only in geometries with more than one hidden layer.}\label{fig:MLP structure}
\end{figure}

The structure of a MLP $\phi$ is shown in Figure~\ref{fig:MLP structure}.

The model is trained using the Adam optimizer~\cite{kingma2017adam} with
$\beta_1 = 0.9$ and $\beta_2 = 0.999$.

The loss function used during the training phase has the following form:
\begin{equation}
    \mathcal{L} = \mathcal{L}^\text{LCAS} + \alpha \cdot \mathcal{L}^\text{Mass}\,,
    \label{eq:loss}
\end{equation}
where $\mathcal{L}^{\text{LCAS (Mass)}}$ is a cross-entropy
loss~\cite{cross-entropy} quantifying the difference between the model's output
and the training dataset for the LCAS (mass hypotheses) predictions, while $\alpha$
is a parameter controlling the relative importance of the two terms, and is set
to $1$ in the default configuration. Various multi-objective optimization
methods were tried, but this arbitrary choice lead to the best results.

\begin{figure}[htb]
    \centering
    \subfloat[]{\includegraphics[width=0.45\textwidth]{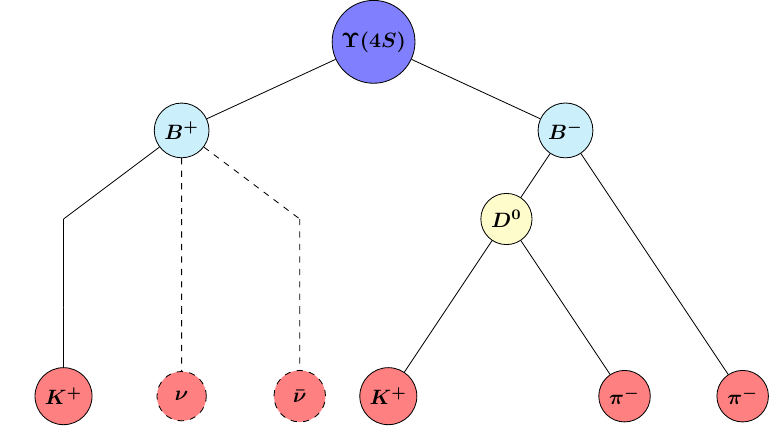}}
    \hspace{1cm}
    \subfloat[]{\includegraphics[width=0.25\textwidth]{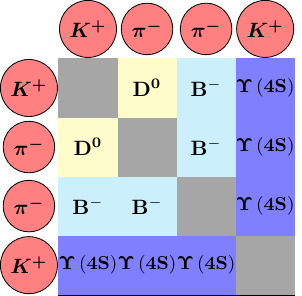}}
    \caption{\label{fig:tree LCA} Schematic representation of a decay tree (a)
    and its LCA matrix representation (b). One can then use the classes defined
    in subsection~\ref{subsec:problem_statement} to write the LCAS matrix. The
    diagonal is represented as empty since the common ancestor of a particle
    with itself is not defined.}
\end{figure}

In addition to the cross-entropy loss, two other metrics are used to evaluate
the performances of the model. The \textbf{perfectLCA} metric is defined as the
fraction of events in the dataset with a perfectly reconstructed LCAS matrix.
An example of such a matrix and the associated decay tree is shown in
Figure~\ref{fig:tree LCA}. 
The \textbf{perfectMasses} metric is defined as the fraction of events in the
dataset where all the final state particles have the correct mass hypothesis
assigned. The \textbf{perfectEvent} metric combines the two: it evaluates the
fraction of events with a perfectly reconstructed LCAS matrix and where all the
final state particles have the correct mass hypothesis assigned.

Moreover, an additional requirements called \textbf{validTree} is defined.
Despite not being used during the training phase, it is useful to reject badly
reconstructed events when applying the model on data. In general, an arbitrary
matrix does not describe a valid LCAS tree representation. Consider exchanging
a $\pi^-$ and $K^+$ in Figure~\ref{fig:tree LCA}. This operation
corresponds to altering \textit{two} entries in the LCAS matrix. If only one is
changed, the obtained matrix does not describe a coherent tree structure anymore. 

\begin{figure}[htb]
    \centering
    \begin{minipage}[b]{0.45\textwidth}
        \includegraphics[width=\textwidth]{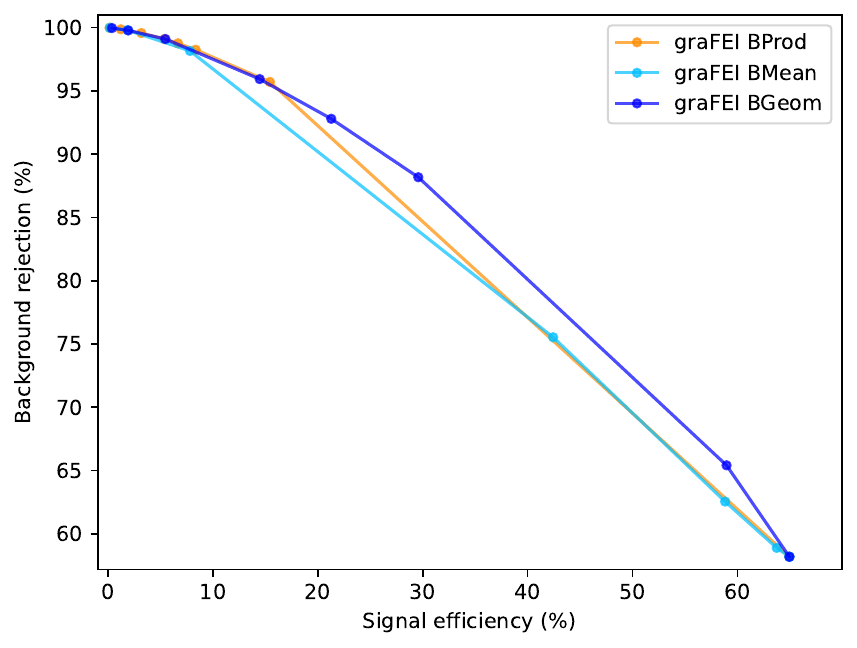}
    \end{minipage}
    \begin{minipage}[b]{0.45\textwidth}
        \caption{Signal efficiency VS background rejection for the three definitions of graFEI $B$ probability.\label{fig:Bprobs}}
    \end{minipage}
\end{figure}

In order to perform a selection on data, a numerical quantity derived from the
\textsc{graFEI} is needed. This quantity should be interpreted as a likelihood for a
given LCA matrix to describe a $B$ decay, or a ``$B$ probability''. Three
quantities were investigated:
\begin{itemize}
    \item $\texttt{BMean} \equiv - \frac{1}{I}\sum_{i=1}^I \log{\frac{\exp{(x_{i, y_i})}}{\sum_{c=1}^C \exp{(x_{i,c})}}}\,$;
    \item $\texttt{BProd} \equiv - \prod_{i=1}^I \log{\frac{\exp{(x_{i, y_i})}}{\sum_{c=1}^C \exp{(x_{i,c})}}}\,$;
    \item $\texttt{BGeom} \equiv \sqrt[I]{\texttt{BProd}}\,$;
\end{itemize}
 After studying the performances of the three quantities, the
\texttt{BGeom} was chosen as the most effective, as one can see from figure \ref{fig:Bprobs}.

\section{Proof of concept: the \texorpdfstring{$B^+ \to K^+ \nu \bar \nu$}{B+ to K+ nu nubar} decay}
\subsection{State of the art}
\begin{figure}[htb]
    \centering
    \subfloat[\label{fig:state of the art}]{\includegraphics[width=0.45\textwidth]{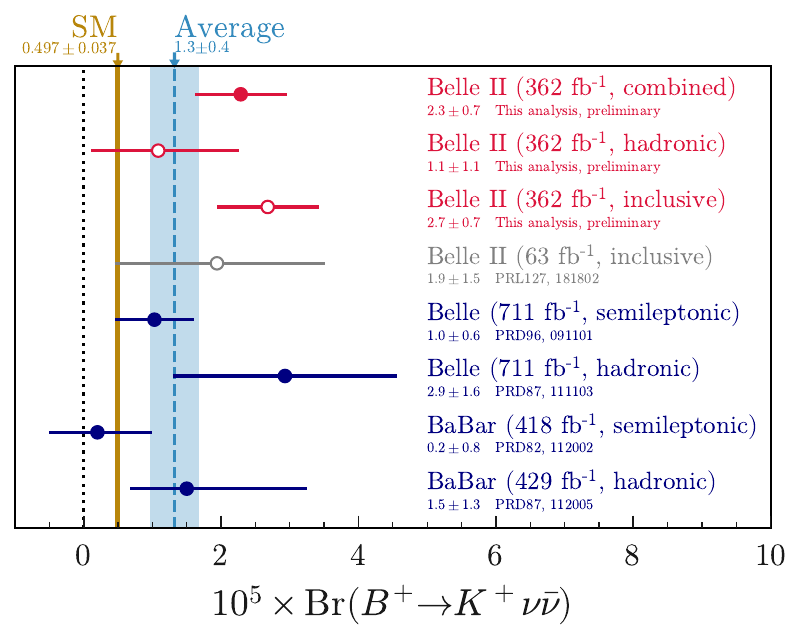}}
    \hspace{0.5cm}
    \subfloat[\label{fig:nll hta ita}]{\includegraphics[width=0.45\textwidth]{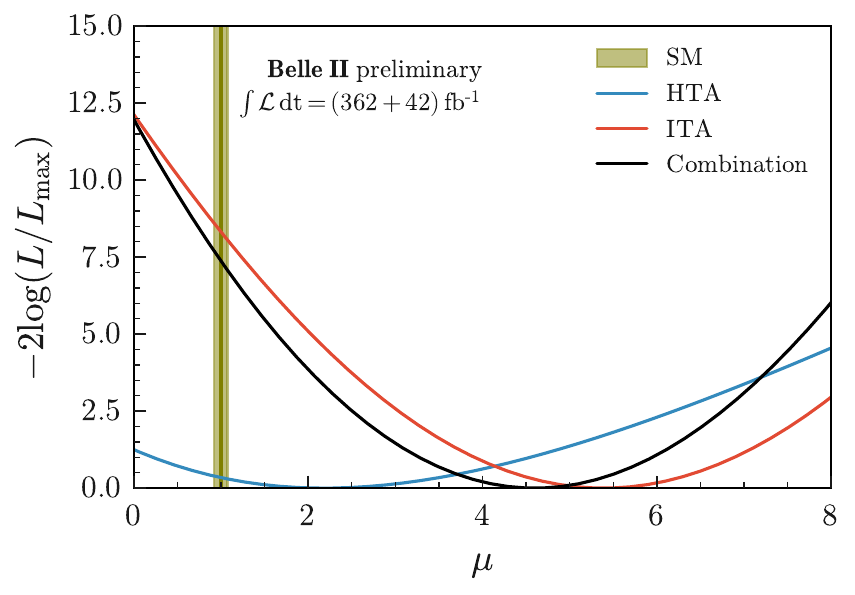}}
    \caption{State of the art of the $B^+ \to K^+ \nu \bar \nu$ search.    
    \protect\subref{fig:state of the art} shows the branching ratio of the decay
    for the inclusive reconstruction and \textsc{FEI} analyses done at Belle~II
    compared to the previous results.  \protect\subref{fig:nll hta ita} shows
    the negative log-likelihood of the \textsc{FEI} and inclusive analyses with
    respect to the \textit{signal strength} $\mu$, the ratio between the
    experimental and the standard model prediction of the branching ratio. \cite{Belle-II:2023esi}}
\end{figure}

This tool was applied to the search for the $B^+ \to K^+ \nu \bar \nu$ decay at
Belle~II.  As previously mentionned, two techniques, inclusive reconstruction
and \textsc{FEI}, were used for this search, leading to two distinct analyses.
The inclusive reconstruction technique is called \textit{Inclusive
Tagged Analysis} (ITA) and the \textsc{FEI} technique is called \textit{Hadronic
Tagged Analysis} (HTA).  Those analyses lead, combined, to a hint for a
deviation from the standard model prediction at $2.7\sigma$, as shown by figures
\ref{fig:state of the art} and \ref{fig:nll hta ita}.

\subsection{Analysis strategy}
In order to compare the performances of the \textsc{graFEI} model with the
\textsc{FEI} and inclusive analyses, we need to define a similar analysis strategy, named \textit{GTA}:

\begin{itemize}
    \item \textbf{Reconstruction and preselection}: We select events where the
    LCAS contains the correct topology. We then ask that the event have a
    $\texttt{BGeom}>0.2$, before applying a cut on the kaon likelihood of the
    signal kaon $\left(>0.9\right)$ and a cut on the cosine of the angle between the thrust
    axis of the signal $B$ meson and the thrust axis of the rest of the event $\left(<0.9\right)$.
    \item \textbf{Training of a binary classifier}: The remaining background is
    further rejected using a BDT implemented with XGBoost. The BDT is trained
    using 10\% of the dataset, further divided into an 80\% for training and
    20\% for testing, and applied to the remaining 90\% of the simulation (the
    loss in efficiency is corrected accordingly in the final result). The BDT is
    trained using 21 input variables.
    \item \textbf{Signal region cut}: The BDT distribution is subsequently
    flattened on signal events. A signal-enhanced region is defined by applying
    a cut on $\text{BDT} > 0.8$. The signal efficiency for this cut is about
    18\%.  
\end{itemize}

\subsection{Comparison}

We start by defining the signal purity $\mathcal{P}^{\text{sig}}$ as the ratio between the number of
signal events over the number of signal and background events after the whole
strategy.  The total efficiency of the technique, $\varepsilon$, will simply be the product of
the efficiency at each step as described in the previous subsection.  The
results are summarized in the table \ref{tab:comparison}. Even with a simply optimised analysis flow,
the GTA shows a better purity than the ITA and
an efficiency 6 times better that the HTA.  This result is very promising and
shows that the \textsc{graFEI} model can be used to significantly improve the
performance of the \textsc{FEI} technique while having a better signal purity than the
inclusive technique.

\begin{table}[htb]
    \hspace{1.5cm}
    \begin{minipage}[r]{0.5\textwidth}
        \begin{tabular}{|c|c|c|}
            \hline
            & $\varepsilon [\%] $ &  $\mathcal{P}^{\text{sig}}[\%]$  \\
            \hline
            HTA  & 0.4 & 3.5\\ \hline
            ITA & 8 & 0.8\\ \hline
            \textbf{GTA} &  \textbf{2.7} & \textbf{1.3} \\ \hline
        \end{tabular}
    \end{minipage}
    \hspace{-1.5cm}
    \begin{minipage}{0.45\textwidth}
        \caption{Comparison of the efficiency $\varepsilon$ and signal purity
        $\mathcal{P}^{\text{sig}}$ for the \textsc{FEI} (HTA), inclusive (ITA)
        and \textsc{graFEI} (GTA) analyses.\label{tab:comparison}}
    \end{minipage}
\end{table}

\begin{figure}[htb]
    \centering
    \begin{minipage}[b]{0.45\textwidth}
        \includegraphics[width=\textwidth]{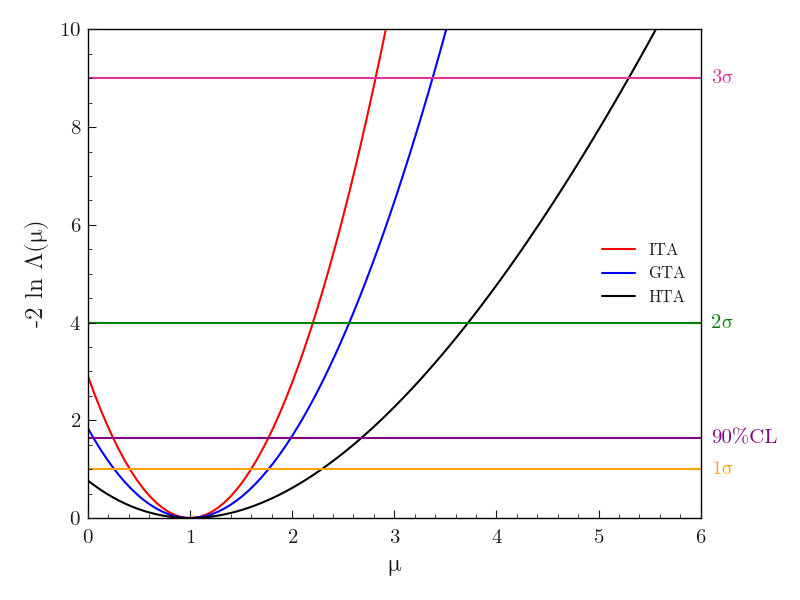}
    \end{minipage}
    \begin{minipage}[b]{0.45\textwidth}
        \caption{Negative log-likelihood of the \textsc{FEI}, inclusive and
        \textsc{graFEI} analyses with respect to the signal strength
        $\mu$.\label{fig:final comparison}}
    \end{minipage}
\end{figure}

We also did the comparison of the likelihood for each analysis, similarly to
figure \ref{fig:nll hta ita}, but without considering any systematic uncertainty
and normalizing the signal strength to 1 to have a purely statistical comparison
of the methods, as one can see in figure \ref{fig:final comparison}. The GTA is
two times better than the HTA, but 20\% worse than the ITA.  This could however
change with further optimisation of the analysis flow and the addition of the
systematic uncertainties.

\section{Future of the \textsc{graFEI}}

With this work, we demonstrated that the \textsc{graFEI} model is a powerful
tool for the inclusive reconstruction of $B$ meson decays at Belle~II. The model
is capable of reconstructing the $B^+ \to K^+ \nu \bar \nu$ decay without any
prior assumption on the decay topology, leading to a higher efficiency than the
previous exclusive algorithm.  Moreover, the knowledge of the decay topology, as
infered by the \textsc{graFEI}, increase the signal purity with respect to the
previous purely inclusive reconstruction technique. This overall underlines the
ability of the model to classify rare signal events among numerous background
events, similarly proved by the \textsc{DFEI} algorithm of the LHCb
collaboration \cite{GarciaPardinas2023}.  Nevertheless, the analysis strategy
can still be improved, especially the binary classifier. Combined with a higher
statistics, we expect the \textsc{graFEI} to be a powerful tool for the search
of rare decays at Belle~II.

\section*{Acknowledgements}
This work of the Interdisciplinary Thematic Institute QMat, as part of the ITI 2021-2028
program of the University of Strasbourg, CNRS and Inserm, was supported by IdEx Unistra (ANR 10 IDEX 0002),
and by SFRI STRAT'US project (ANR 20 SFRI 0012) and EUR QMAT ANR-17-EURE-0024 under the framework of the French Investments for the Future Program,
the Institut National de Physique Nucléaire et de Physique des Particules (IN2P3) du CNRS (France), 
the Centre de Calcul de l’IN2P3 (CC-IN2P3), the French Agence Nationale de la Recherche (ANR) under 
Grant ANR-21-CE31-0009 (Project FIDDLE).
\printbibliography

\end{document}


<div id='footer'><table width='100